\documentclass[twocolumn,twoside,slac_two]{revtex4}
\usepackage{graphicx}
\usepackage{fancyhdr}
\renewcommand{\headrulewidth}{0pt}
\renewcommand{\footrulewidth}{0pt}

\begin{document}

\title{Supporting the GLAST User Community}

\author{David L. Band}
\author{GLAST Science Support Center}
\affiliation{Code 661, NASA/GSFC, Greenbelt, MD  20854}

\begin{abstract}
The Gamma-ray Large Area Space Telescope (GLAST) Science
Support Center (GSSC) is the scientific community's
interface with GLAST.  The GSSC will provide data, analysis
software and documentation.  In addition, the GSSC will
administer the guest investigator program for NASA HQ.
Consequently, the GSSC will provide proposal preparation
tools to assist proposers in assessing the feasibility of
observing sources of interest.
\end{abstract}

\maketitle

\thispagestyle{fancy}

\section{Introduction}
The GLAST Science Support Center (GSSC) will support the
scientific community's analysis of the GLAST data.  Here we
describe the data analysis opportunities that will be
available to the community through the GSSC's services.
However, first we describe the mission and its
capabilities.
\section{GLAST Mission Overview}
GLAST is an international and multi-agency space mission
that will study the cosmos in the 10~keV--300~GeV energy
range.  The launch is currently scheduled for the end of
May, 2007, on a Delta II into low earth orbit (565~km).

The main instrument, the Large Area Telescope (LAT), will
have an effective area ($>$8000~cm$^2$), angular resolution
($<$3.5$^\circ$ at 100 MeV, $<$0.15$^\circ$ at $>$10~GeV),
field-of-view ($>$2 sr), and deadtime ($<$100 $\mu$s) that
will provide a factor of 30 or more advance in sensitivity
compared to previous missions, as well the capability for
studying transient phenomena.  The field-of-view (FOV) of
$>$2~sr is the effective area integrated over the sky
divided by the peak effective area; usable observations can
be performed up to $\sim$70$^\circ$ from the LAT's axis.
The GLAST Burst Monitor (GBM) will have a FOV larger than
that of the LAT and will provide spectral coverage of
gamma-ray bursts extending from the LAT's lower limit down
to 10~keV. Although pointed observations will be possible,
the observatory will most likely scan the sky continuously
because of the LAT's large FOV; this survey mode is planned
for at least GLAST's first year.  In survey mode the
spacecraft's axis is rocked perpendicular to the orbital
plane once per orbit, achieving near-uniform exposure every
two orbits.  Thus in survey mode the spacecraft's pointing
changes continuously.

\begin{figure}
\includegraphics[width=80mm]{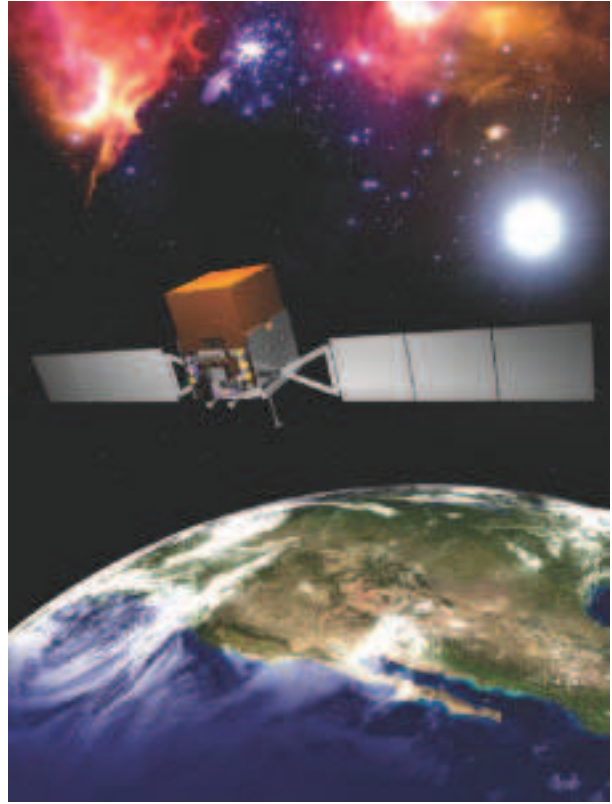}
\caption{GLAST in Orbit.  Figure courtesy of SASS, the
spacecraft contractor.}
\end{figure}
\subsection{The LAT}
A product of a NASA/Department of Energy/international
collaboration, the LAT builds on the success of the
Energetic Gamma Ray Experiment Telescope (EGRET) on the
{\it Compton Gamma Ray Observatory (CGRO)}.  The PI is
P.~Michelson (Stanford) and the instrument is managed at
the Stanford Linear Accelerator Center (SLAC).  The LAT
will be a pair conversion telescope (see Figure~2): gamma
rays will pair-produce in tungsten foils; silicon strip
detectors will track the resulting pairs; the resulting
particle shower will deposit energy in a CsI calorimeter;
and an anticoincidence detector will veto the large flux of
charged particles that will also be incident on the LAT.
The anticoincidence detector will be segmented to eliminate
the self-vetoing at high energy that plagued EGRET. The
LAT's outside dimension will be 1.8m$\times$1.8m$\times$1m,
and it will weigh $\sim$3000~kg.

\begin{figure}
\includegraphics[width=80mm]{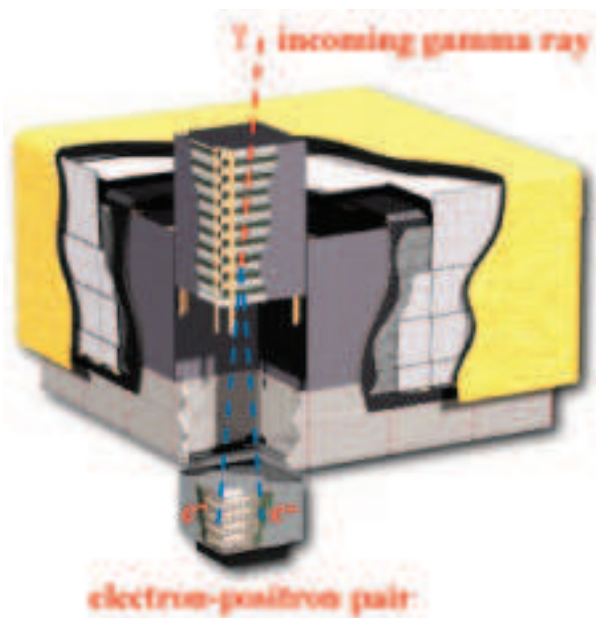}
\caption{Structure of the LAT---the gamma ray interacts in
a tungsten foil which is housed in the assembly that is
shifted up in this cutaway illustration. The resulting
electron-positron pair passes through the silicon strip
detectors interspersed with the tungsten foils, and the CsI
calorimeter (shifted down in this figure) below the
tungsten foils and silicon strip detectors.  Charged
particles are vetoed by the anticoincidence detector, the
light grey tiles under the yellow thermal blanket.}
\end{figure}

Astrophysical photons will be only a small fraction of all
the events the LAT will detect, most of which will be
charged particles.  Therefore, event filtering on board
will reduce the $\sim$4~kHz detected event rate to
$\sim$300~Hz that will be telemetered to the ground; ground
processing will identify the true $\sim$2--3~Hz photon
rate.

The tracks of the particle shower produced by a photon or
cosmic ray interacting in the LAT will be analyzed by the
LAT Instrument and Science Operations Center (LISOC)
located at SLAC, resulting in a determination of whether
the event was a photon, and a characterization of the
event's arrival time, direction and energy.  The list of
detected photons is the primary dataset that will be used
for astrophysical analysis.
\subsection{The GBM}
A descendant of {\it CGRO}'s Burst And Transient Source
Experiment (BATSE), the GBM will detect gamma-ray bursts
and extend GLAST's burst spectral sensitivity to the
$<$10~keV to $>$25~MeV band.  The PI is C.~Meegan (MSFC)
and the co-PI is G.~Lichti (MPE).  Consisting of
12~NaI({\it Tl}) (10--1000 keV) and 2~BGO (1--25~MeV)
detectors, the GBM will monitor $>$8~sr of the sky,
including the LAT's FOV. Bursts will be localized to
$9^\circ$ (1$\sigma$, brightest 40\% of the bursts) by
comparing the rates in different detectors.  The GBM will
trigger if the rates in $\ge$2 detectors increase
simultaneously by $\ge 5.5\sigma$. The trigger will use a
variety of energy bands and time windows.

From the GBM's telemetry the GBM Instrument Operations
Center (GIOC) will produce `continuous' and burst data
products.  The primary continuous data are two sets of
rates from all the GBM's detectors with differing temporal
and spectral resolution, regardless of whether a burst was
detected.  The primary burst data are lists of the counts
in each detector from the period of the burst.  Both the
continuous and burst data products include calibration
data, catalogs, and other ancillary data.
\subsection{GLAST Science}
GLAST will study a wide range of energetic astrophysical
phenomena.  In many cases GLAST will extend EGRET's
pioneering observations, but we anticipate new phenomena
will be revealed.  The references provide an entree into
the relevant literature.

Active Galactic Nuclei (AGN)\cite{blazar,AGN} are extremely
energetic sources observed at the center of some galaxies;
AGN are believed to be powered by accretion onto
super-massive black holes (10$^6$--10$^{10}$ solar masses).
EGRET detected $\sim$70~AGN whose gamma-ray emission is
believed to be radiated by relativistic jets pointed in our
direction. We expect GLAST will detect several thousand
AGN.

Gamma-ray bursts (GRBs)\cite{GRB} are short, very bright
flashes of gamma-rays followed by fading afterglows at
lower energies (radio, optical and X-ray).  EGRET saw 45
high energy gamma-ray photons (in total) from several GRBs,
including an 18~GeV photon 75~minutes after a
burst.\cite{GRB_GEV} GLAST will determine whether there are
additional spectral and temporal high energy components.
The synergy between the GBM and the LAT is crucial for
providing spectra over 7~energy decades and for detecting
bursts over a large FOV. GLAST will repoint autonomously
toward strong bursts, keeping them within the LAT FOV for
$\sim$5~hours (except when the burst location is occulted
by the earth). In addition, GLAST will inform the ground
within less than 10 seconds that a burst has occurred.

The shocks in supernova remnants\cite{SNR,SNR_TEV}
accelerate particles that may then radiate gamma-rays.
Since supernova remnants are thought to be the origin of
the primary cosmic ray population below $10^{15}$~eV,
gamma-ray observations may increase our understanding of
the cosmic ray phenomenon and propagation.

Pulsars\cite{pulsar} are spinning magnetized neutron stars
that emit gamma rays when young.  EGRET identified five
pulsars, but may have detected a number of others as point
sources without discovering their pulsations.  GLAST
observations should distinguish between two competing
explanations of pulsar high energy emission:  the outer
gap\cite{outer_gap} and polar cap models.\cite{polar_cap}

Gamma rays interact with lower energy photons, producing
electron-positron pairs.  Consequently gamma rays
originating at cosmological distances are attenuated while
propagating through the optical-UV radiation field produced
by stars between the source and us.  The energy-dependent
attenuation depends on the density and evolution of the
radiation field.  GLAST will observe cutoffs in the spectra
of AGN and GRBs from which the infrared radiation density
can be measured.\cite{optuv}

The particles that constitute the cosmological dark matter
may annihilate in a cuspy halo around the Galactic Center,
producing a spectral feature that GLAST might
detect.\cite{particle}

The interaction of cosmic rays with the interstellar medium
and inverse Compton emission by the cosmic rays' electron
component result in a diffuse Galactic
emission\cite{background_gal} on top of isotropic diffuse
extragalactic emission thought to be the sum of unresolved
AGN and other components.\cite{background_ext}  While this
diffuse emission is a background complicating the detection
of point sources, it is scientifically interesting.

As a result of GLAST's great increase in sensitivity over
previous missions such as EGRET, we expect to discover new
source classes, and to find that previous classes consist
of subclasses.  GLAST should allow the identification of
the 172~unidentified strong sources in the 3rd EGRET
catalog (out of the 271~detected
sources).\cite{pulsar,catalog,unid_gould}
\section{GLAST Data Policy}
During the first year of the mission, LAT data are
proprietary to the instrument team, although information on
detected transients and $\sim$20 selected sources will be
made public as soon as possible.  During this first year
the LAT team will calibrate their instrument and undertake
an all-sky survey that will result in a point source
catalog. The catalog will be updated in subsequent years.

A month after the end of the first year, these LAT data
will become publicly available. Starting the second year,
all subsequent science data acquired by the mission will be
in the public domain within 24 hours without a proprietary
data period.

GBM data, particularly from bursts, will become publicly
available from the beginning of the mission.

Full details on the GLAST Data Policy will be included in a
public document and will be posted at
http://glast.gsfc.nasa.gov/ssc/data/Data\_Policy.html.
\section{User Support by the GSSC}
The GSSC was established at Goddard Space Flight Center
(GSFC) to take advantage of the synergy with the Office of
General Investigator Programs (OGIP) that runs similar
organizations supporting the {\it RXTE}, Swift, {\it
XMM-Newton}, {\it Integral}, and {\it Astro-E2} missions.
The GSSC therefore draws upon the user support expertise
and infrastructure within OGIP. In particular, OGIP also
houses the High Energy Astrophysics Science Archive
Research Center (HEASARC), NASA's archive for high energy
astrophysics missions.  The HEASARC maintains a common data
storage and analysis environment for high energy missions,
and will be the ultimate archive of GLAST data after the
mission ends and the GSSC is disbanded.  By developing its
database and software systems within the HEASARC
environment, the GSSC insures that the GLAST data will be
easily accessible to the scientific community during and
after the GLAST mission.  In particular, the GLAST Standard
Analysis Environment (SAE) is not yet another analysis
system but adds GLAST-specific tools to the HEASARC's
HEADas software system.  The HEASARC will be the GLAST
data's final archive.

The GSSC has different roles before and after the
observatory's launch:

{\bf Before Launch—--}The GSSC  will educate the user
community about the mission's capabilities through posters
and talks at scientific conferences, special workshops and
tutorial sessions providing hands-on experience with
simulated GLAST data and by maintaining an up-to-date
website with the current status and information about the
GLAST mission.

{\bf During the Mission—--}The GSSC's website will provide
updates about the mission's status, serve as a gateway to
the data, tools, instrument response functions (IRFs) and
documentation, and include a help desk and Frequently Asked
Questions (FAQ) section of the GSSC website. The GSSC will
also host conferences and workshops to provide education
and experience with the GLAST Science Software and a forum
for users to report on GLAST's scientific results.
\section{Guest Investigator Program}
The GLAST mission will support a Guest Investigator (GI)
program that the GSSC will administer for NASA
Headquarters.  The GI program will include a GLAST Fellows
program.  The program will be part of NASA's Research
Opportunities in Space and Earth Science (ROSES), and will
consist of yearly cycles.  The GI program will provide the
opportunity for scientists anywhere in the world to propose
GLAST observations and for investigators at US institutions
to receive funding for their GLAST-related research.

For the mission's first year (the first GI cycle), GIs may
not propose GLAST pointed observations.  During this first
year the LAT team will post information on bright
transients and $\sim$20 selected sources.  GIs may request
funding for multiwavelength observations, support projects
(e.g., developing new analysis methods) and GLAST-related
theoretical research during the first cycle.

During the subsequent yearly cycles, GIs may also request
pointed observations or special instrument modes as part of
their proposal, if scientifically justifiable.  However,
continued surveying of the sky will probably be the most
efficient method of accumulating exposure for the largest
number of sources, and we anticipate that most observing
programs will be satisfied by survey mode.  During this
phase of the mission, all data will be available to the
public from the GSSC's website.

To assist scientists prepare GI proposals, the GSSC will
provide a set of tools for planning observations and
submitting proposals. The proposal planning tools will
include an exposure and sensitivity calculator as well as
observation simulation tools to assist potential GIs assess
the feasibility of observing their desired targets (see
http://glast.gsfc.nasa.gov/ssc/proposals/Proposal\_ \hfill
Tools.html).  These tools will simulate the spacecraft's
orbit and the instruments' observations with varying levels
of fidelity.  For example, an online detectability
calculator will use orbit-averaged exposure accumulation
rates and tables for sources with power law spectra, while
the user will be referred to the simulation tools within
the SAE, the analysis system that will be used to analyze
actual data, for more sophisticated calculations.

Target of Opportunity (TOO) requests will be submitted
through an interface similar to the one used for submitting
GI proposals.
\section{The GSSC Within the GLAST Ground System}
Although the specifics are not directly relevant to the
user community, the GSSC has an important role within the
ground system as the advocate for the community's
scientific goals.  The observations proposed successfully
through the GI program are converted into first an annual,
and then a weekly, science timeline by the GSSC. To ensure
that instrument operations do not disturb the science
timeline, the two instrument operation centers---the LISOC
and GIOC (see \S 2 above)---route instrument commands and
software uploads through the GSSC, which schedules their
implementation.  The GSSC provides the weekly science
timeline as well as the timelines for the two instruments
to the Mission Operations Center (MOC), which integrates
these timelines with the spacecraft timeline, resulting in
a weekly timeline for the observatory.  The science
timelines are posted on the GSSC website at the different
stages of their development to inform the community of the
observatory's observing plan.

The GSSC also plays a central role in processing TOOs.
Requests will be submitted to the mission through the GSSC
website, and the GSSC will evaluate the feasiblity of the
proposed TOO and its impact on the science timeline.  If
the TOO is approved, the GSSC asks the MOC to implement the
TOO.
\section{Providing Data to the Community}
All public data from the GLAST mission will be available
through the GSSC's website (see
http://glast.gsfc.nasa.gov/ssc); Table~1 lists the data
products that will be available. Much of the data will be
served through the HEASARC's Browse interface (an interface
to all of NASA's high energy astrophysics data from both
current and previous missions---see
http://heasarc.gsfc.nasa.gov/db-perl/W3Browse/
w3browse.pl); the GSSC website will link to this interface.
Those data not available through Browse will be served
directly from the GSSC's website.  The data necessary for
the response functions will be stored in the HEASARC's
CALDB directory structure.  Table~1 indicates whether the
access is through Browse, the GSSC website (labelled simply
`GSSC') or CALDB.

\begin{center}
\begin{table*}
\caption{GLAST DATA PRODUCTS}
\begin{tabular}{l l l}
\hline
\bf {Data Product}       & \bf {Description}                                           & \bf {Access} \\
\hline
 & \qquad \bf{SCIENCE DATA}& \\
\hline
LAT Events                & Full detailed description of events (particle and gamma-rays) reconstructed  & Browse \\
                          & by the LAT                   & \\
\hline
LAT Photons               & LAT events considered to be photons. Includes all the data necessary to & Browse \\
                          & calculate the instrument response functions (IRFs). & \\
\hline
LAT IRFs                  & Data necessary to calculate LAT IRFs                       &  CALDB \\
\hline
LAT Burst Catalog         & Catalog of burst information derived from the LAT          &  Browse \\
\hline
LAT Point Source Catalog  & Detected gamma-ray sources with derived information        & Browse \\
\hline
Interstellar Emission Model& Model for diffuse Galactic and extragalactic gamma-ray emission & GSSC \\
\hline
LAT Transient Data       & Summary information for transient sources GRBs, (solar flares, AGN flares) &  GSSC \\
\hline
GBM CTIME                & For each detector, the counts accumulated every 0.256 s in 8 energy channels& Browse \\
\hline
GBM CSPEC                & For each detector, the counts accumulated every 8.192 s in 128 energy channels& Browse \\
\hline
GBM Calibration          & Tables of fiducial detector response parameters from which the burst-specific  &  GSSC \\
                         & DRMs are calculated & \\
\hline
GBM Time Tagged Events   & Time tagged events from the GBM centered on the time of triggered GRBs &  Browse \\
\hline
GBM Burst DRMs           & DRMs for the burst, one for each significantly different pointing &  Browse \\
\hline
GBM TRIGDAT              & The GBM burst alert data in a single FITS file &  Browse \\
\hline
GBM Background Files     & GBM Background Files &   Browse \\
\hline
GBM Burst Catalog        &  List and characterization of all bursts & Browse \\
\hline
GBM Trigger Catalog      & List and characterization of all GBM triggers  &  Browse \\
\hline
GBM Burst Spectra Catalog& Catalog of deconvolved spectra &  Browse \\
\hline
Pulsar Ephemerides       & Ephemerides of pulsars that might be detectable by GLAST & Browse \\
\hline
GCN Notices and Circulars&  GCN notices and circulars generated by GLAST            & Browse \\
\hline
Accepted GI Proposals    &  Database of Accepted GI Proposals                       &  GSSC \\
\hline
& \qquad \bf{MISSION DATA}  & \\
\hline
Science Timelines         & Long (annual) and short (weekly) term science timelines describing where the &  GSSC \\
                          & spacecraft was or will be pointing & \\
\hline
\quad $\bullet$ Long Term Science   & Planned observing schedule for the current yearly cycle in 1 week periods.  & GSSC \\
\qquad Schedule           & Updated as needed & \\
\hline
\quad $\bullet$ Preliminary Science & Preliminary detailed one week observing schedule prepared 3 weeks in advance & GSSC \\
\qquad Timelines          &  & \\
\hline
\quad $\bullet$ Final Science Timelines & Final detailed one week observing schedule used to generate commands & GSSC \\
                           & uploaded to the spacecraft.  Generated $\sim$3 days before upload and replaces the & \\
                           & Preliminary Timeline once generated &\\
\hline
\quad $\bullet$ As-flown timeline  & Timeline describing the history of GLAST's pointing & GSSC \\
\hline
LAT pointing and         & LAT orientation and mode at 30 s intervals.  These data are used to calculate & Browse \\
\quad livetime history   & exposures & \\
\hline TOO Data          & Lists of all accepted and executed Target of Opportunity requests and their & GSSC \\
                         & status &\\
\hline
\end{tabular}
\end{table*}
\end{center}

\section{GLAST Standard Analysis Environment (SAE)}
The GSSC will provide a suite of data analysis tools and
libraries for the analysis of GLAST data.  This software is
being developed by the instrument teams with assistance
from the GSSC.  The instrument teams and the scientific
community will all use the SAE suite, which will run on
Windows and different flavors of UNIX platforms, and will
not require the purchase of additional software.  Most of
the SAE will be implemented as FTOOLs, and all will be part
of the HEADas system maintained by the HEASARC.
Consequently the data files input and output from the tools
will be in FITS format. Therefore, the SAE will be an
extension of the data analysis environment familiar to the
high energy astrophysics community.  In addition, we are
developing a GUI interface to run these tools.

The SAE can be divided into a number of analysis areas:

{\bf General Analysis---}The SAE will consist of several
general purpose tools including a data sub-selection tool,
tools to generate source models and extract source
parameters from existing catalogs, and the workhorse of the
GLAST data analysis, the Likelihood tool to perform maximum
likelihood fits of the data with the specified models (see
below) . The suite also provides an event binning tool to
create time, energy and spatially binned data sets and
tools to compute exposure and response matrices.

{\bf GRBs---}The SAE suite will provide several tools to
assist in the study of gamma-ray bursts including tools for
spectral and temporal data analysis and model fitting as
well as tools for generating the necessary response
functions and binning events for analyzing GLAST data with
existing tools such as XSPEC. These tools will be used to
analyze both LAT and GBM data, either individually or
simultaneously.  The GBM team will also provide their
IDL-based burst analysis tool RMFit, to which they will add
the capability to analyze both LAT and GBM data.

{\bf Pulsars---}The SAE suite will include a barycenter
arrival time correction tool, period search and profiling
tools, and a pulsar ephemeris extraction tool to retrieve
pulsar ephemerides from a pulsar database.

{\bf Data Simulation---}The SAE suite also provides an
observation simulator that simulates LAT data based on an
input source model and spacecraft orbit profile.  An orbit
simulation tool is also included.
\section{Why a New Likelihood Tool?}
One of the primary goals in the analysis of LAT data is to
find the location and spectral parameters of high energy
gamma-ray sources.  This information is obtained by
performing maximum likelihood fits of spatial-spectral
models to the data.  A major new likelihood tool is
required for the LAT data because:

The large LAT point spread function at low energy and the
great sensitivity means photons from many point sources
merge.  Analysis is therefore inherently three dimensional:
two spatial and one spectral.

GLAST will usually survey the sky.  Each photon will
therefore have a different direction in instrument
coordinates, and thus its own instrument response.

The LAT instrument response will be a function of many
observables, such as energy, distance from the source, and
angle between the photon direction and the LAT's axis. The
data space of observables will therefore be large but
sparsely populated.

Ideally, the analysis will be unbinned, i.e., using
infinitesimally small bins in the data space, each
containing 0 or 1 photons.  However, the runtime for large
datasets is prohibitively long, and therefore a binned
version is under development.
\section{GLAST Users' Committee}
The GLAST Users' Committee (GUC) has been established; the
chair is J.~Grindlay (Harvard).  The GUC reviews the
mission's support for the scientific community in general,
and the scientific role of the GSSC in specific.
Conversely, the GSSC supports the GUC by facilitating
meetings and gathering information.  The GUC has been
considering issues such as the data policy, the GI program
and the SAE.  Information on the GUC and its membership can
be found on the GSSC's website.
\section{Summary}
The identification and characterization of photons detected
by the LAT is a complex data analysis task that will be
undertaken by the LAT team.  Most investigators will
analyze a simple list of photons characterized by a few
observables such as energy, arrival time, and direction in
both celestial and instrument coordinates.  Similarly, the
GBM's primary burst data will be a list of counts detected
in the different GBM detectors.  Although the astrophysical
data analysis problem is simply posed, the techniques
necessary to extract the maximum information from these
data are sophisticated and computer-intensive. The goal is
to create an analysis system that uses advanced techniques
but is easy to learn and use.  The GSSC will provide the
scientific community with these photon lists as well as
ancillary data, with analysis software, and with the
expertise to analyze the data.  In addition, the GSSC will
support the GI program that will provide investigators with
the possibility of requesting pointed observations and with
the funding necessary to carry out their research.

\end{document}